\documentclass[twocolumn,trackchanges]{aastex631}
\usepackage{float}

\shorttitle{QPOs in Swift J1727.8$-$1613}
\shortauthors{Zhu and Wang}

\graphicspath{{./}{figures/}}
\begin{document}

\title[Energy dependence of QPOs in Swift J1727.8$-$1613 ]{Energy dependence of the low-frequency quasi-periodic oscillations in Swift J1727.8$-$1613}

\author{Haifan Zhu}
\affiliation{Department of Astronomy, School of Physics and Technology, Wuhan University, Wuhan 430072, China}
\author{Wei Wang}
\affiliation{Department of Astronomy, School of Physics and Technology, Wuhan University, Wuhan 430072, China}

\begin{abstract}
Based on observations from the Insight-Hard X-ray Modulation Telescope (\textit{Insight-}HXMT), an analysis of Type-C quasi-periodic oscillations (QPOs) observed during the outburst of the new black hole candidate Swift J1727.8-1613 in 2023 was conducted. This analysis scrutinized the QPO's evolution throughout the outburst, particularly noting its rapid frequency escalation during two flare events. Utilizing the energy range covered by \textit{Insight-}-HXMT, a dependency of the QPO frequency on energy was observed. Below approximately 3 Hz, minimal variations in frequency with energy were noted, whereas clear variations with photon energy were observed when it exceeded approximately 3 Hz. Additionally, a sharp drop in the rate of change was observed when the frequency exceeded approximately 8 Hz. This behavior, similar to several previously reported sources, suggests the presence of a common underlying physical mechanism. Moreover, the QPO rms-frequency relationship can be explained by the Lense-Thirring precession model. The relationship between rms-energy and phase lag with frequency suggests the black hole system as a high-inclination source.

\end{abstract}


\section{Introduction} \label{sec:intro}
Most black hole X-ray binaries (BHXBs) are transient sources, characterized by alternating periods of dormancy marked by low flux or quiescent states, and outbursts featuring high flux or active states over their evolutionary trajectory. These binary systems primarily inhabit a quiescent phase, typified by low X-ray luminosity, with their emissions remaining undetectable until the onset of outbursts. During these episodes, X-ray emissions experience a significant surge, often raising by several orders of magnitude, and display distinct spectral states. This intermittent transition from prolonged periods of faint luminosity during quiescence to sporadic outbursts is a defining characteristic of low-mass black hole X-ray binaries \citep{remillard2006}.

During the outburst rise phase, the system transitions from a quiescent state to increased luminosity, characterized by strong variability and spectra dominated by nonthermal emissions in the low-hard state (LHS) \citep{basak2016spectral}.
As the outburst progresses, the source transitions to the high-soft state (HSS), characterized by the dominance of thermal disc emission in the X-ray spectrum \citep{sharma2018spectral}. An intermediate state (IMS) exists between these two states. It can be further divided into the hard-intermediate state (HIMS) and soft-intermediate state (SIMS). During the IMS, the X-ray spectrum exhibits characteristics of both the LHS and the HSS. At the end of an outburst, the X-ray flux from the black hole X-ray binary typically fades quickly, commonly by 1$\sim$2 orders of magnitude, as the accretion rate decreases, and the system returns to the LHS \citep{homan2005evolution, belloni2016astrophysics}.

Quasi-periodic oscillations (QPOs) are frequently observed during the outburst phase of black hole X-ray binaries, manifesting in the power spectrum as narrow, harmonically related peaks. They can be categorized based on their frequency range into low-frequency QPOs (LFQPOs) and high-frequency QPOs (HFQPOs). Regarding their spectral and timing characteristics, LFQPOs in black hole binaries (BHBs) are further subdivided into Type-A, Type-B, and Type-C categories.
The classification criteria primarily revolve around parameters such as the quality factor $Q = \nu / \Delta \nu$ (where $\nu$ represents the frequency of the QPO and  $\Delta \nu$ represents the full width at half maximum; FWHM), fractional root-mean-square (rms) variability, noise component, and phase lag \citep{casella2005abc,motta2015geometrical,motta2016quasi,ingram2019}.

Studying LFQPOs is crucial for enhancing our comprehension of the accretion process surrounding black holes, even though their origin remains a topic of debate. The majority of models employed to elucidate LFQPOs revolve around either the geometric configuration of the accretion disk or its intrinsic characteristics.
The Lense-Thirring precession model \citep{stella1997lense,stella1999correlations,schnittman2006precessing} stands out as the most viable approach, positing that LFQPOs arise from the relativistic precession of either an inner hot accretion flow \citep{ingram2009low,ingram2011physical} or a miniature jet \citep{ma2021discovery, ma2023detailed}.
Investigating the energy-dependent timing characteristics of QPOs, including the fractional rms amplitude, centroid frequency, and the phase lags among various energy bands\citep{qu2010energy,motta2015geometrical,ingram2016quasi,van2016inclination,zhang2017evolution,karpouzas2021variable}, can provide insights into the underlying radiative mechanisms driving QPOs.

Although the Lense-Thirring (L-T) model has successfully explained numerous observed phenomena, recent research results have contradicted this model.
\cite{marcel2021can} suggest that the Lense-Thirring mechanism overlooks realistic properties of the accretion flow and propose that Type-C QPOs may result from the extreme accretion speed and disc-driven ejections.
 In a recent study by \cite{nathan2022phase}, an intriguing finding emerged concerning GRS 1915+105, revealing an exceptionally minute inner disc radius ($r_{in}\sim 1.4~ GM/c^2$), a revelation at odds with the predictions made by existing models. This diminutive disc radius poses a challenge to the efficacy of the L-T precession model, as it fails to account for the observed QPO frequency associated with such a minuscule radius.

Intrinsic models include resonant oscillations in a property of
the accretion flow such as accretion rate, pressure, or electron
temperature. For instance, the Accretion Ejection Instability (AEI) model \citep{Tagger1999,varniere2002accretion}, as described by \cite{Tagger1999}, represents a spiral wave instability occurring within the density and scale height of a slender disk that is penetrated by a robust vertical (poloidal) magnetic field. Another model, known as the Propagating Oscillatory Shock (POS) model, is based on interactions between the disk and corona \citep{chakrabarti1993smoothed,Molteni1996, chakrabarti2008evolution}.

The new X-ray transient Swift J1727.8$-$1613, initially recognized as GRB 230824A, was first discovered with \textit{Swift}/BAT on August 24, 2023 \citep{Page2023}.
The subsequent observation revealed a rapid increase in flux from the source, identifying it as a new galactic X-ray transient \citep{Kennea2023GCN,negoro2023maxi}.
Subsequent observations in optical \citep{castro2023optical}, radio \citep{miller2023vla}, and X-ray \citep{o2023nicer} also indicate the source was a black hole candidate.
Type-C QPOs have been observed in X-ray data \citep{palmer2023swift, draghis2023preliminary,sunyaev2023integral,katoch2023detection,bollemeijer2023nicer,mereminskiy2023hard}, and additionally, the Imaging X-ray Polarimetry Explorer (IXPE) has also conducted observations of this source \citep{dovciak2023ixpe,dovciak2023ixpeb}.

\cite{sanchez2024evidence} analyzed optical spectroscopy data to estimate the orbital period to be approximately 7.6 hours. Employing various empirical methods, they derived a distance to the source of
$d=2.7\pm 0.3~\rm kpc$, corresponding to a Galactic plane elevation of
$z=0.48\pm0.05~\rm kpc$ and this source exhibited complex signatures of optical inflows and outflows throughout its discovery outburst.

\cite{veledina2023discovery} presented the first detection of X-ray polarization from Swift J1727.8$-$1613 using the IXPE. They proposed that the hot corona, responsible for emitting the majority of the detected X-rays, exhibits an elongated rather than spherical morphology. Furthermore, the alignment of the X-ray polarization angle with that observed in submillimeter wavelengths suggests that the corona's elongation is orthogonal to the jet direction \citep{ingram2023tracking}. \cite{zhao2024first} presented the first polarimetric analysis of QPOs in Swift J1727.8$-$1613  utilizing IXPE data.
Subsequent phase-resolved analysis of the QPO, utilizing the Hilbert–Huang transform technique, unveiled a significant modulation of the photon index in relation to the QPO phase. However, discrepancies emerged as the polarization degree (PD) and polarization angle (PA) displayed no such modulation, contrasting the anticipated outcomes from the L-T precession model. This incongruity underscores the need for additional theoretical investigations to reconcile with the observed phenomena.

Spectral analysis was conducted using simultaneous observations from \textit{Insight-}HXMT, and the Neutron Star Interior Composition Explorer (\textit{NICER}), and the Nuclear Spectroscopic Telescope Array (\textit{NuSTAR}) during the transition of this source towards the hard intermediate state \citep{peng2024nicer}. The study inferred a spin of approximately 0.98 and an orbital inclination of approximately 40 degrees for this source. Based on observations from \textit{Insight-}HXMT,  temporal and spectral analysis of this source indicated that it is a high-inclination, high-spin source \citep{WeiYu2024, chatterjee2024insight}.

In this paper, we present the energy dependence of the  Type-C QPOs in Swift J1727.8$-$1613 using \textit{Insight-}HXMT. Section \ref{obs} describes the observations and data reduction methods. Section \ref{result} provides our QPO results. Sections \ref{Discussions}
and \ref{conclu} present the discussion and conclusions.

\section{OBSERVATIONS AND DATA REDUCTION}
\label{obs}
\subsection{Observations}

From August 25, 2023, to October 4, 2023, Swift J1727.8$-$1613 underwent observations by the \textit{Insight}-HXMT. This instrument comprises three detectors operating across distinct energy bands: High Energy (HE) detectors covering 20.0 to 250.0 keV with a temporal resolution of 4 $\rm \mu s$ \citep{liu2020High}, Medium Energy (ME) detectors spanning 5.0 to 30.0 keV with a temporal resolution of 240 $\rm \mu s$ \citep{cao2020medium}, and Low Energy (LE) detectors ranging from 1.0 to 15.0 keV with a temporal resolution of 1 $\rm ms$ \citep{chen2020low}. The corresponding detector areas are 5100 $\rm cm^2$, 952 $\rm cm^2$, and 384 $\rm cm^2$, respectively.

The extraction and analysis of data were conducted utilizing Version 2.04 of the \textit{Insight-}HXMT Data Analysis Software (HXMTDAS)\footnote{\url{http://hxmten.ihep.ac.cn/software.jhtml}}. We established our good time intervals based on specific criteria: a pointing offset angle below 0.04$^\circ$, an earth elevation angle exceeding 10$^\circ$, a geomagnetic cutoff rigidity surpassing 8$^\circ$, and the exclusion of data within 300 seconds of passing through the South Atlantic Anomaly (SAA). Light curves were generated using the $helcgen$, $melcgen$, and $lelcgen$ tools in HXMTDAS. Background estimation was performed using HEBKGMAP, MEBKGMAP, and LEBKGMAP, further refined using $lcmath$ to eliminate estimated background noise. Our analysis focused on energy ranges of 1--10 keV (LE), 10--20 keV (ME), and 20--100 keV (HE). This outlined procedure was applied consistently to process all available open data for Swift J1727.8$-$1613 across the entire observation period.

\subsection{Analysis methods}

To investigate the variability signals, we employed \textit{powspec} from HEASOFT to calculate the power density spectra (PDS) for
each observation, using a time interval of 64 seconds and a corresponding time resolution of 1/128 seconds. The resulting
PDS was re-binned in frequency space using a geometric factor of 1.05. We normalized the PDS to units of $\rm rms^2/Hz$ and
subtracted Poisson noise following the methods described in \cite{belloni1990atlas, miyamoto1991x}. Subsequently, we
identified observations with clear QPO signals. Some observations with lower count rates were
excluded due to poor quality. Ultimately, 96 observations with robust QPO signals were selected.

For the fitting process, we utilized multiple Lorentz functions to fit the profiles of the broadband noise (BBN), QPO, and their harmonics in the PDS. This approach allowed us to extract the fundamental parameters of the QPO. To determine the fractional root mean square (rms) of the QPO, we calculated
\begin{equation}
\rm rms_{QPO}=\sqrt{R}\times\frac{S+B}{S},
\end{equation}
where S represents the source count rate, R represents the normalization of the QPO's Lorentzian component, and B represents the background count rate \citep{bu2015correlations}. As our primary focus was on low-frequency QPOs, the frequency range selected for the fitting was 0 to 30 Hz.

To examine the energy-dependent properties of the QPO, we partitioned the \textit{Insight}-HXMT data into various energy ranges: 1--3 keV, 3--5 keV, 5--7 keV, 7--10 keV, 10--13 keV, 13--16 keV, 16--20 keV, 20--40 keV, 40--60 keV and 60--100 keV. We calculated the cross spectra of the light curves obtained from the low-energy and high-energy bands. The phase-lags between the different energy bands were computed by taking the 1--3 keV band as the reference. Positive lags, or hard photons lagging behind soft photons, were referred to as hard lags. To determine the necessary phase-lags and other parameters, we employed the $stingray$ \footnote{\url{https://github.com/StingraySoftware/stingray/}} package \citep{huppenkothen2019stingray}. We calculated the phase lags of the QPOs over a frequency range centered on the QPO centroid frequency and dispersed over its FWHM. We estimated the uncertainties of the parameters using Markov Chain Monte Carlo (MCMC) simulations with 200 walkers and a length of $10^5$ steps. The uncertainties reported in this paper are given at the $90 \%$ confidence level. Additionally, we segmented the LE data into two energy bands, 2--4 keV and 4--10 keV, to calculate the hardness ratio.

\section{RESULTS}
\label{result}
\subsection{Light curves and Type-C QPOs of Swift J1727.8-1613 }
\begin{figure}
    \includegraphics[width=\columnwidth]{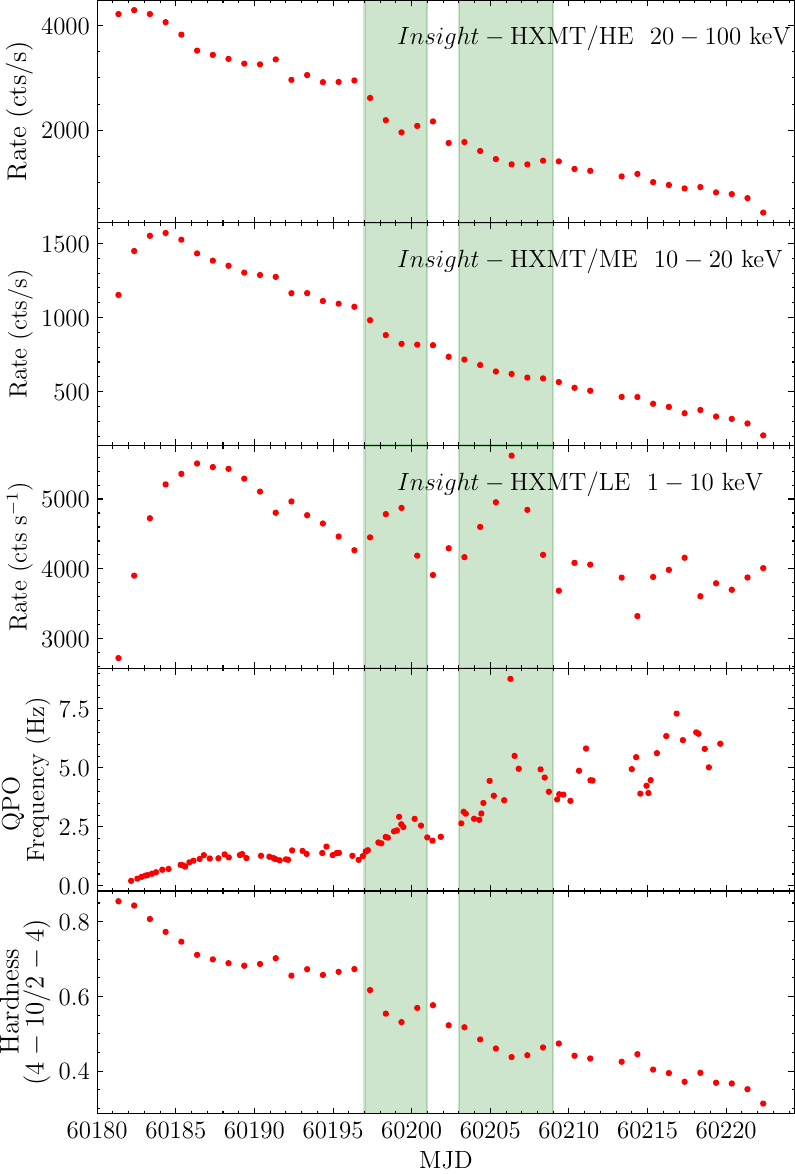}

    \caption{The Swift J1727.8$-$1613 light curves are presented in the following sequence,
    proceeding from top to bottom: \textit{Insight}-HXMT HE (20--100 keV) data, \textit{Insight}-HXMT ME (10--20 keV) data,
    \textit{Insight}-HXMT LE (1--10 keV) data, the QPO frequency in the LE energy band.,
    and hardness ratios derived from \textit{Insight}-HXMT data using  (4--10 keV)/(2--4 keV). The green shaded region represents the periods of abrupt increases and decreases in count rate within the LE energy band. }
    \label{figure1}
\end{figure}

\begin{figure}
    \includegraphics[width=\columnwidth]{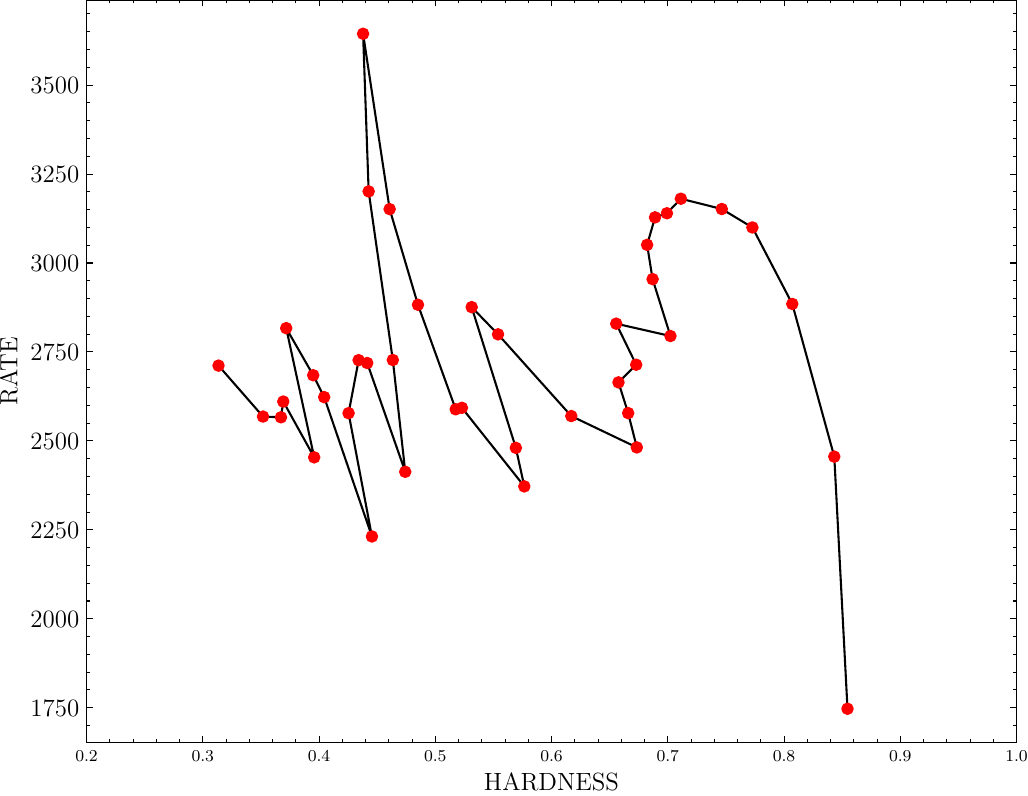}   
    \caption{The hardness-intensity diagram (HID) of Swift J1727.8-1613. On the horizontal axis, hardness ratios are represented 
    using data acquired from \textit{Insight}-HXMT LE (4--10 keV/2--4 keV). The vertical axis corresponds to the count rate of \textit{Insight}-HXMT LE (1--10 keV). 
}
    \label{figure2}
   \end{figure}

Figure~\ref{figure1} presents the background-subtracted light curves from \textit{Insight}-HXMT,
along with the frequency of the fundamental type-C QPO and hardness ratios.
The hardness-intensity diagram (HID) of Swift J1727.8-1613 is illustrated in Figure~\ref{figure2}.

\begin{figure*}[]
  \includegraphics[width=0.3\textwidth]{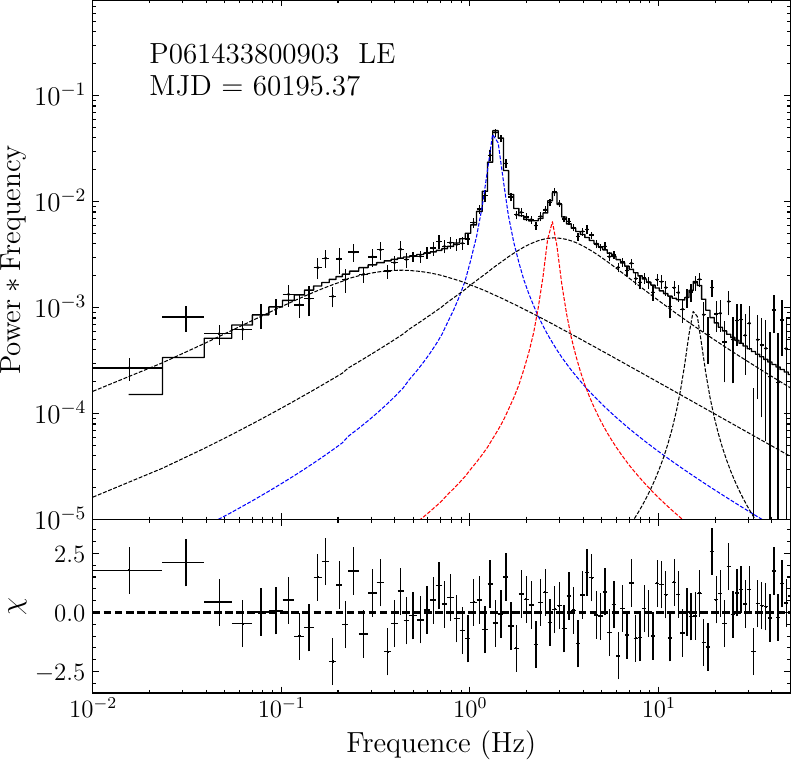}
  \includegraphics[width=0.3\textwidth]{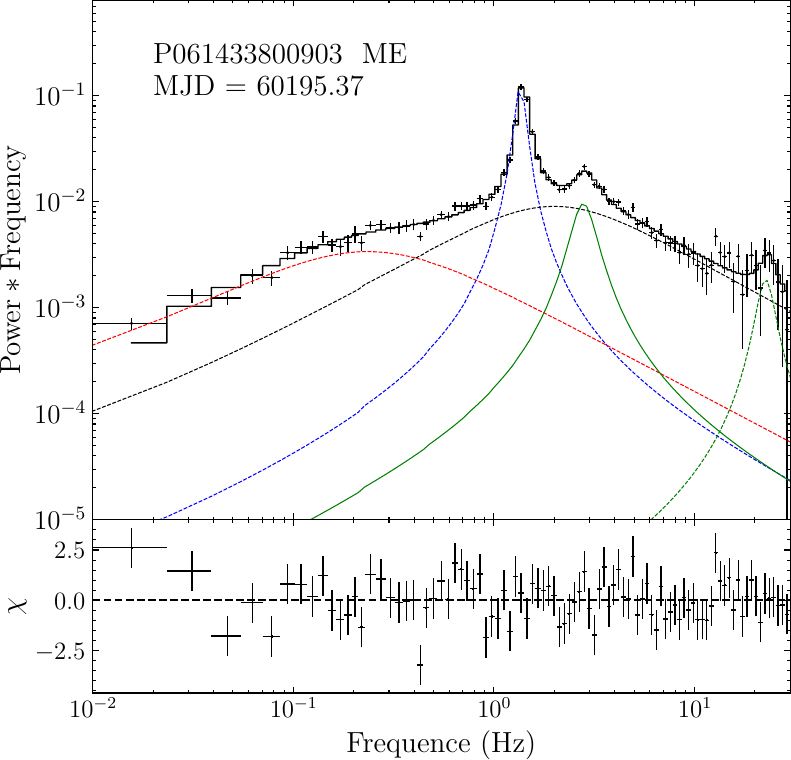}
  \includegraphics[width=0.3\textwidth]{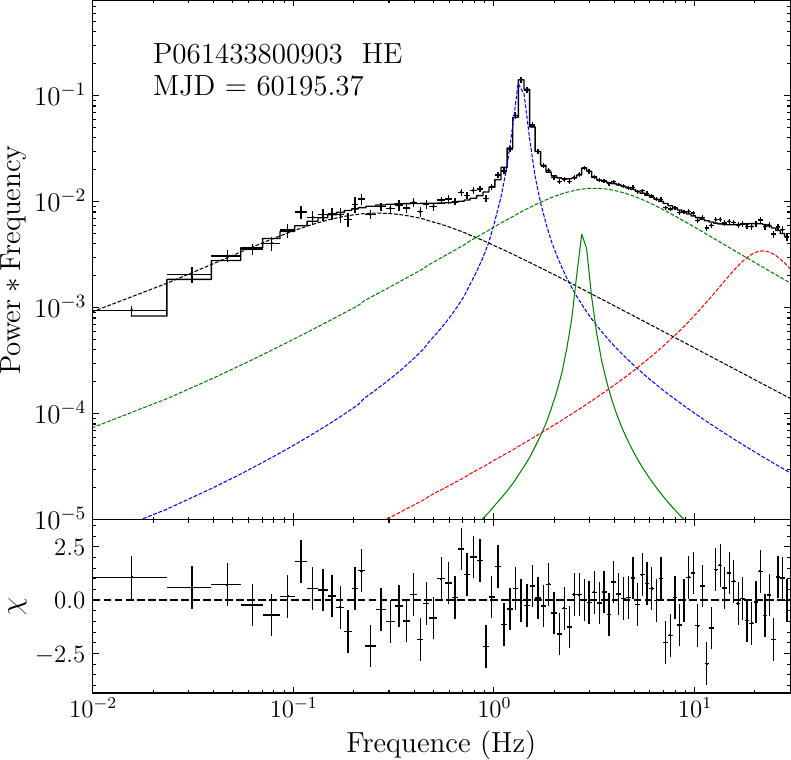}
  \caption{The representative fitting results of the QPO signals were obtained using data from 
  \textit{Insight-}HXMT LE, ME, HE. 
  The ObsID and MJD are labeled in the respective figures. We utilize distinct colors and line styles to 
  differentiate the Lorentz components.} 
  \label{hxmtpds}
\end{figure*}

In Figure~\ref{hxmtpds}, we present the example of the PDS obtained from the three detectors of the \textit{Insight}-HXMT, accompanied by their respective model fitting results. In the observations we selected,
 harmonics are prevalent in both the LE and ME energy bands, whereas  harmonic occurrences in the HE energy band are sparse,
 appearing only in a minority of observations. The frequency uncertainty obtained from the fitting are generally about two orders of magnitude smaller than the frequencies themselves.

Initially, the count rate in the HE energy band peaks at the beginning of the observation,
reaching approximately 4000 $\rm cts ~s^{-1}$, followed by a gradual decrease over time.
In the ME energy band, the count rate starts from ~1150 $\rm cts ~s^{-1}$ and gradually decreases after reaching its peak.
Conversely, in the LE band, the count rate exhibits an increase from 2700 $\rm cts~s^{-1}$ at the beginning,
reaching a peak of 5600 $\rm cts~s^{-1}$ on MJD 60186. Subsequently, it gradually declines. However, a distinctive pattern emerges in the LE energy band starting from MJD 60197,
characterized by a sudden increase in count rate followed by a rapid decline.
Another notable change occurs at MJD 60203, where the count rate reaches approximately 5700 $\rm cts~s^{-1}$.
These two events are marked with green shaded regions in Figure \ref{figure1}.
Importantly, there are no anomalous changes in the count rates in the HE and ME bands,
which continue to decrease gradually over time. We refer to the two distinct changes of LE count rates occurring between MJD 60197 to 60201 and MJD 60203 to 60209 as "flare-1" and "flare-2," respectively.

The fourth plot in Figure~\ref{figure1} illustrates the fundamental frequency of the QPOs corresponding to the 96 selected observations, essentially covering the entire outburst process.
In the initial stages, the frequency of the QPO signal gradually increases from 0.2 Hz to 1.3 Hz over time.
In the subsequent periods of flare-1 and flare-2, the frequency of QPO and the flux within the LE energy range exhibit analogous trends, while the flux in the ME and HE domains experiences a sustained decline. This phenomenon
is also reported by \cite{bollemeijer2023nicer} using NICER data. The observed increases in the QPO frequency
seem to intricately correspond with flaring in the soft X-ray flux, as observed by both NICER and MAXI.
Meanwhile, the hard X-ray flux measured by MAXI and Swift/BAT continues its descent.  In Figure~\ref{figure2}, we depict the HID of Swift J1727.8$-$1613,
which aligns with the evolutionary trajectory obtained in previous studies \citep{zhao2024first,peng2024nicer,nandi2024discovery}.

\begin{figure}
  \includegraphics[width=\columnwidth]{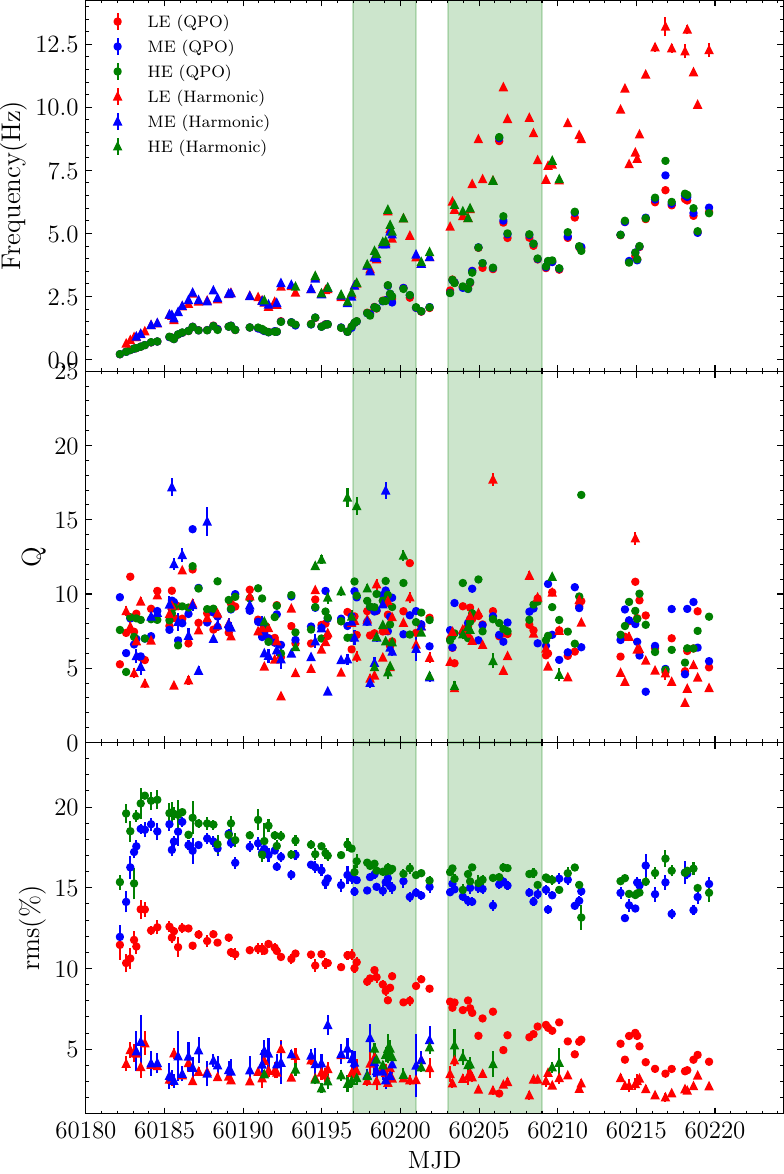}
  
  \caption{The Type-C QPOs and it's harmonics  frequency, Q factor, and fractional rms from the LE energy band as functions of time for Swift J1727.8–1613. 
}
  \label{timeallqpo}
 \end{figure}

We plotted all the fitting results in Figure ~\ref{timeallqpo}, where circular symbols represent QPOs, triangular symbols denote harmonics, 
the red denotes LE data, blue denotes ME data, and green denotes HE data. The Q values of the QPO primarily concentrate between 5 and 10, 
whereas the rms values, with the exception of a few isolated observations, consistently exceed 5. The Q values of the harmonics also 
predominantly lie within this range, while the rms values are considerably smaller, mainly around 5$\%$. Furthermore, throughout the entire observation period, 
the rms values of the harmonics did not exhibit significant variations, maintaining a relatively stable state. The study of harmonics is not a primary focus of this work. we have merely presented the fitting results. 

\subsection{The QPO frequency variation with energy} \label{subsec:tables}

 In the uppermost graph of Figure~\ref{timeallqpo}, a notable convergence of LE, ME, and HE data points is evident at lower 
 frequencies. However, as frequency increases, subtle deviations among the three datasets emerge, indicating energy-dependent 
 fluctuations within the QPO phenomenon at higher frequencies. As previously outlined, we partitioned the 1--100 keV energy 
 band of \textit{Insight}-HXMT into 10 sub-bands, conducting fitting analyses for the QPO phenomena observed within each sub-band. 
 This systematic investigation unveiled more pronounced variations.

\begin{figure*} 
  \includegraphics[width=1\textwidth]{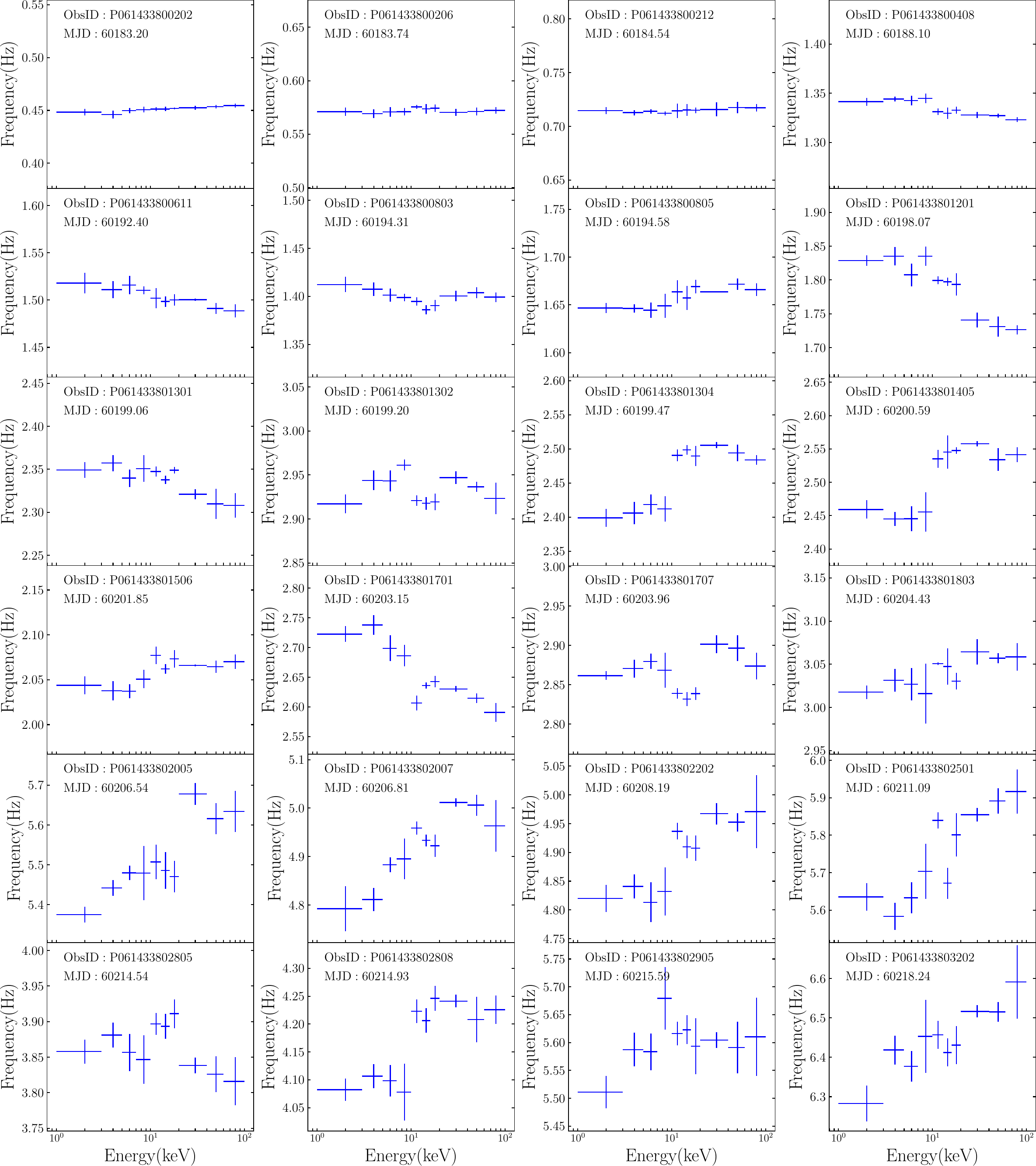}

  \caption{The QPO frequency varies with energy, and the observational IDs and timestamps are annotated above each subplot. } 
  \label{selecthz}
\end{figure*}

 In delineating the relationship between QPO frequency and energy, diverse patterns of change have come to light. 
 In Figure~\ref{selecthz}, we have meticulously selected exemplars that typify these heterogeneous associations. 
 Within this figure, it is observed that the dependency of frequency on energy does not immediately manifest in the early stages; 
 however, it gains prominence with escalating frequency, showcasing a spectrum of relationships,  
 These encompass both positive and negative correlations with energy, as well as instances of initial ascent succeeded 
 by descent. In some observations, such as ObsID: P061433801304 and P061433801305, abrupt changes occur between LE and ME. These sudden changes happen during flare periods and similar transitions are observed in other sources\citep{li2013energy,li2013energyH1743} as well, which may reflect certain changes in the source.

\begin{figure}
  \includegraphics[width=\columnwidth]{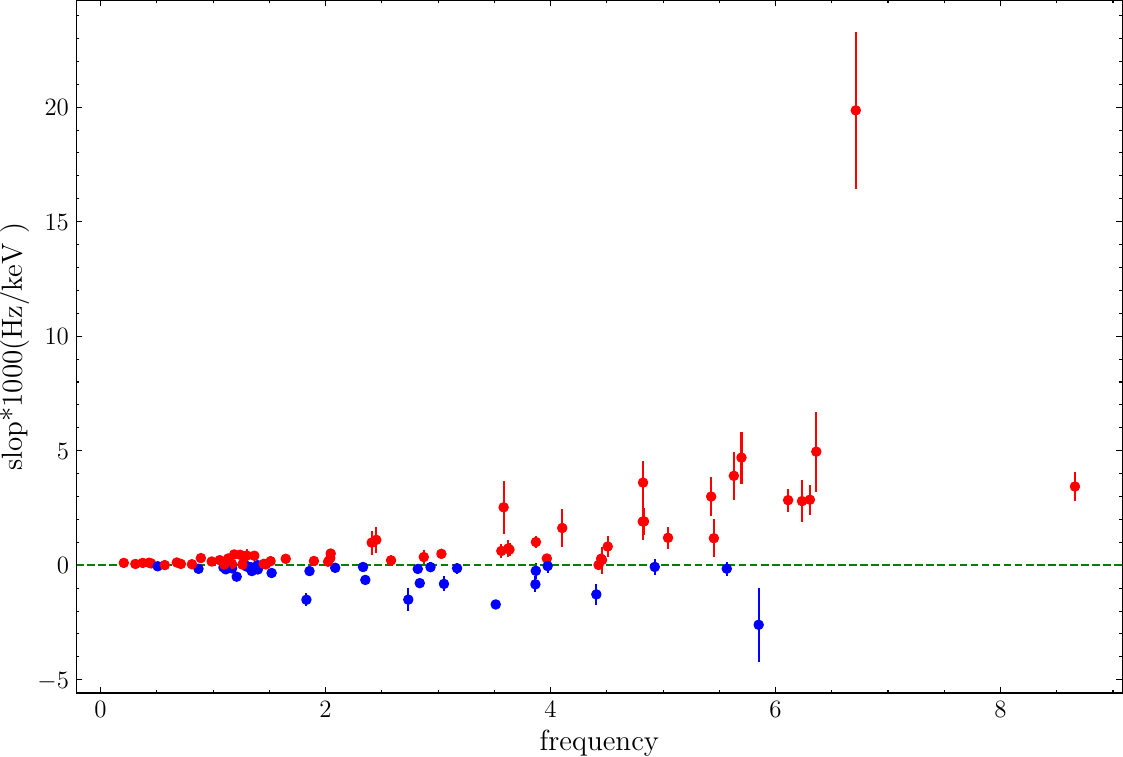}
  
  \caption{The slopes of the QPO frequency-energy relations versus frequency. The red dots in the graph represent positive 
  slopes obtained from the fitting, while the blue dots represent negative slopes. The green dashed line indicates a slope 
  of zero, and the frequency values are based on the fitting results from LE..
}
  \label{slopes}
 \end{figure}

 Like previous studies on the frequency-energy relationship\citep{yan2012systematic,li2013energy,li2013energyH1743}, in order to perform a quantitative analysis of this trend, we employed the least-squares method to estimate the linear correlation between the frequency and photon energy. 
Subsequently illustrated the progression of the fitted slopes as a function of QPO frequency in Figure~\ref{slopes}. Upon examination of this graph, it becomes evident that when the QPO frequency is 
 relatively low ($<$~3 Hz), the slopes tend to remain in close proximity to zero. Among the points demonstrating a 
 positive correlation, the slope experiences a gradual increase with frequency. At higher frequencies ($>$~3 Hz), the 
 slope undergoes a rapid increase, and upon surpassing 8 Hz, it reverts to a comparatively lower level. Additionally, in Figure~\ref{slopes}, we have indicated negative slopes with blue color. The negative slope values do not change much with frequency; they mostly remain within a relatively stable numerical range. There are no negative slopes observed after frequencies greater than 6 Hz.

\subsection{The variations of QPO rms and phase lags \label{subsec:rms}}

\begin{figure}
  \includegraphics[width=\columnwidth]{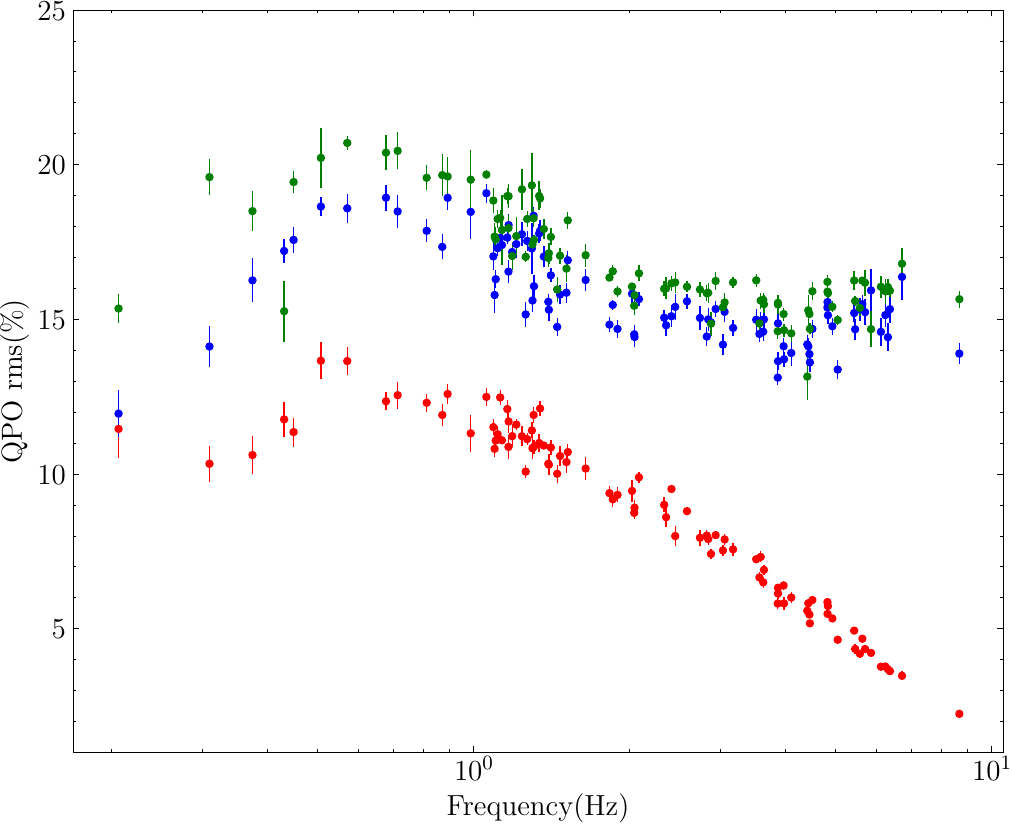}
  
  \caption{The QPO rms relations versus energy, with the red color representing LE data, 
  the blue color signifying ME data, and the green color denoting HE data.
}
  \label{rmsmu}
 \end{figure}

The relationship between the QPO rms and observation time is illustrated in the bottom subgraph of Figure~\ref{timeallqpo}. 
Additionally, the relationship between the QPO rms and frequency is depicted in Figure~\ref{rmsmu}. In the LE, ME, and HE energy bands, when the QPO 
frequency is below 0.5 Hz, the QPO rms values exhibit a slow increase. Subsequently, in all three energy bands, the QPO rms 
values decrease with the rise in frequency. However, in the ME and HE energy bands, the QPO rms displays a slight increase at 
approximately 1.5 Hz and beyond, while the QPO rms in the LE energy band continues to decline. 

\begin{figure*} 
  \includegraphics[width=1\textwidth]{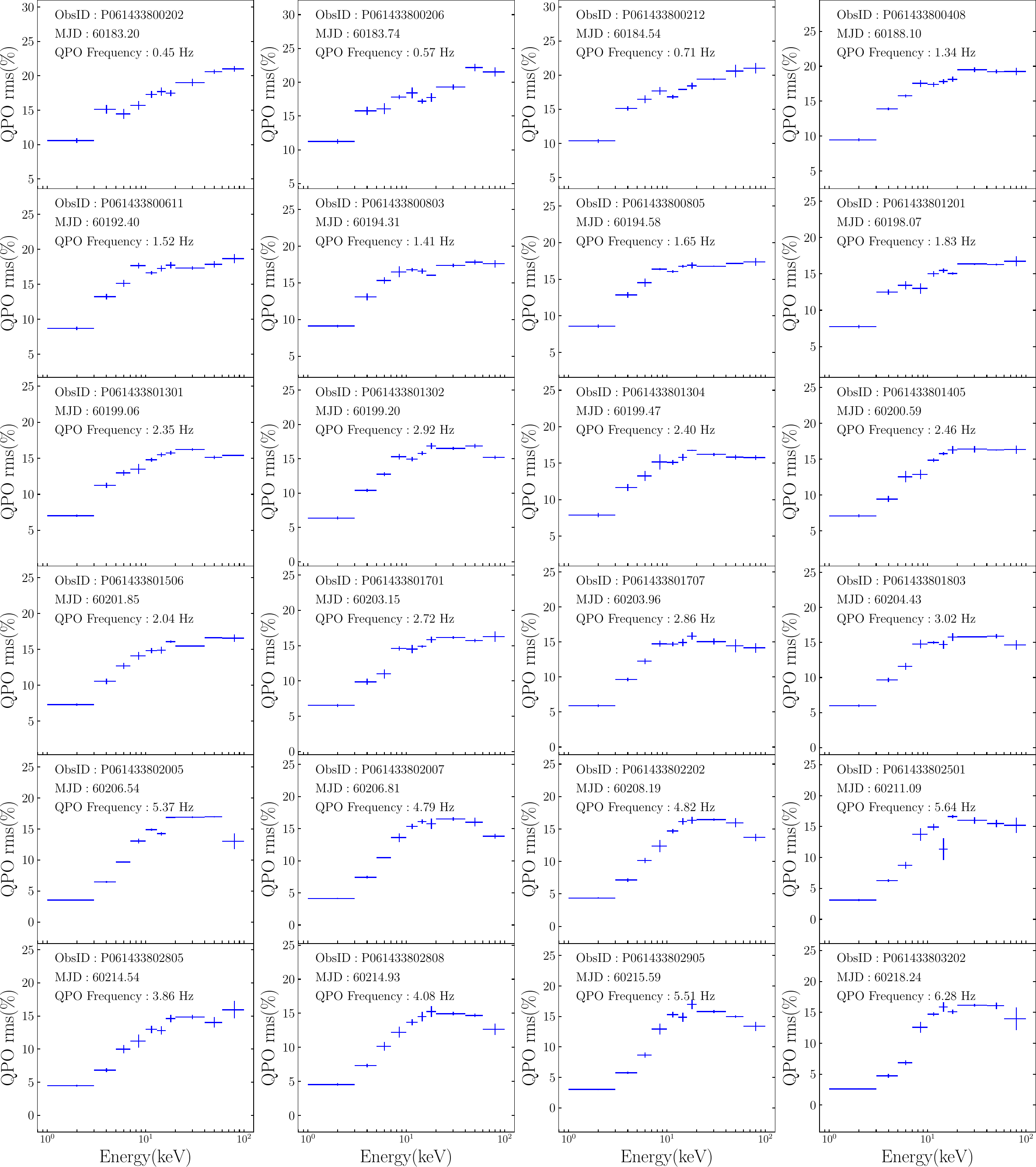}
  
  \caption{The QPO rms as a function of energy, with observational IDs, time,
   and the corresponding frequencies of the Low Energy (LE) bands, indicated above each sub-plot.} 
  \label{rmsenergy}
\end{figure*}

Figure~\ref{rmsenergy} presents a selection of example graphs illustrating the relationship between the QPO rms and energy. The graph
reveals a consistent variation trend, characterized by a distinct increase in the QPO rms in proportion to energy up to 10 keV, 
succeeded by an essentially flat pattern beyond the 10 keV threshold. 

\begin{figure}
  \includegraphics[width=\columnwidth]{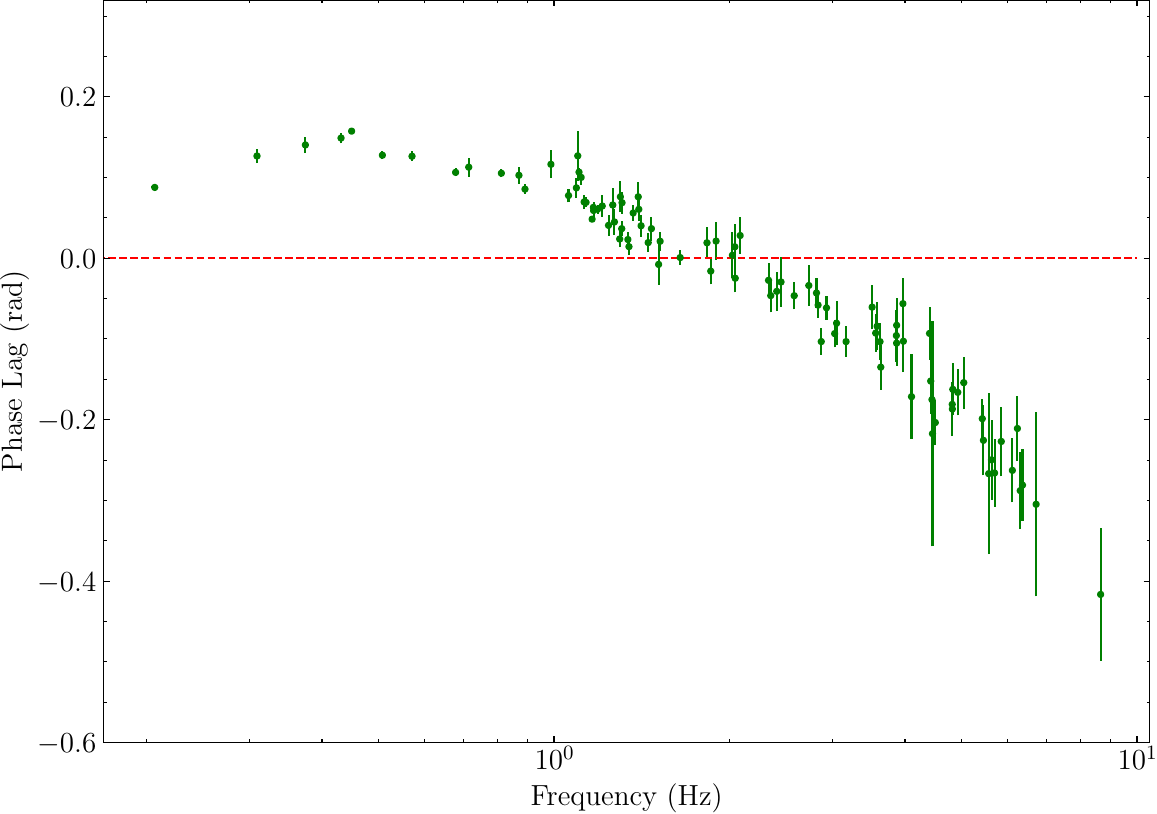}
  
  \caption{The QPO phase lag relations versus frequency for Swift J1727.8$-$1613. The phase lag values are derived from the calculation 
  between two energy bands, 3--5 keV and 1--3 keV, utilizing the 1--3 keV band as a reference. 
  The red dashed line in the graph indicates the point at which the phase lag is equal to zero. 
}
  \label{phasefre}
 \end{figure}

Figure~\ref{phasefre} illustrates the variation of the phase lag with frequency, at lower frequencies, 
the phase lag assumes positive values, approaching zero near a frequency of approximately 1.5 Hz, 
after which the phase lag becomes negative as the frequency increases.  

\section{Discussions} \label{Discussions}

\subsection{Energy-dependent QPO frequency}
A comprehensive analysis of the QPO phenomena exhibited by Swift J1727.8$-$1613 during its outburst has been 
conducted based on the data from \textit{Insight}-HXMT. Throughout the observation, the source exhibited a typical BH XRBs outburst pattern in HID; however, two instances of flaring (flare-1 and flare-2) were observed in 
the LE data, while the ME and HE data exhibited a continuous decrease in flux. 
These two flaring events could potentially be attributed to sudden increases in the mass accretion rate surrounding the black hole, 
However, a more detailed analysis would require the results from spectral fitting, 
which is currently not feasible due to the prominent photon pileup effect observed in \textit{Insight-}HXMT
spectrum \citep{peng2024nicer}.  

Upon fitting the QPO properties within the 10 energy bands we defined, it was discovered that the relationship between QPO frequency and energy exhibits different morphologies depending on the QPO frequency. There are primarily three morphologies observed: flat, frequency increasing with energy, and frequency decreasing with energy, as illustrated in Figure~\ref{selecthz}.   
By analyzing the slope fitting results, depicted in Figure~\ref{slopes}, we observe that below approximately 3 Hz, the QPO frequency remained nearly constant with energy. However, for frequencies exceeding 3 Hz, the frequency exhibited a rapid rise with energy, peaking at a maximum slope of approximately $0.0198\pm 0.0034$ around 6.7 Hz, then declining to about $0.0034\pm 0.0006$ at the maximum frequency of approximately 8.6 Hz. Furthermore, it is also observed that there are many instances of negative slopes present.

Such occurrences have previously been observed in only a few sources. 
For GRS 1915+105, the frequency-energy slope increased from a negative value to approximately 0.1 
as its fundamental frequency increased from 0.7 to 6.3 Hz, followed by a sharp decrease as its frequency 
increased from 6.3 to 8 Hz \citep{qu2010energy,yan2012systematic}. For XTE J1550-564, the centroid frequency of the 
fundamental QPO exhibited varying relationships with photon energy, 
showing minimal variation below ~0.4--0.8 Hz, a clear increase above ~3.3 Hz, and a sharp decline in the rising rate beyond 
~8.5 Hz\citep{li2013energy}. For H1743-322, the QPO frequency exhibited a similar relationship, the slope rapidly increased 
after approximately 3 Hz, peaking around 7.5 Hz, then swiftly declined to lower levels at about 8.5 Hz \citep{li2013energyH1743}.   
For MAXI J1535$-$571 \citep{huang2018}, the presentation of their findings reveals various patterns: the first pattern illustrates an increase in frequencies with increasing photon energy, followed by a decrease after reaching approximately 10 keV. The second pattern suggests that the frequency remains nearly constant, irrespective of photon energy. In the remaining patterns, a consistent trend of monotonically increasing frequencies with photon energy is apparent. 

This phenomenon has been explained by \cite{van2016probing}  in terms of differential Lense-Thirring precession, 
where different parts of the inner accretion flow precess at varying rates. 
However, this explanation is problematic, as it fails to account for the negative correlation observed in the frequency-energy relationship.  
For instance, the blue data points in our Figure~\ref{slopes} exhibit a negative correlation in the frequency-energy relationship. 
Given that the characteristic frequencies and material temperature in the vicinity of a compact object typically 
decrease with increasing distance from the central object, this explanation would necessitate that the outer, 
cooler regions of the accretion flow precess at a faster rate than the inner, hotter regions. 

For MAXI J1535$-$571, \cite{huang2018} suggests that the turnover in the relationship at high energies (E$>$10 keV) indicates the prominence of the reflection bump at those energies. It is anticipated that the reflected spectrum will be predominantly influenced by photons emitted from the outer-part flow of the inner accretion flow, resulting in a relatively lower precession frequency observed in the reflected spectrum. Considering the morphology of the frequency-energy spectrum, our findings suggest that there is no evident turnover in the relationship at energy levels exceeding 10 keV. 

\cite{mendez2024unveiling} introduced an innovative method to analyze an observation of GRS 1915+105, 
wherein \cite{qu2010energy} identified an increase in the QPO frequency with energy. 
\cite{mendez2024unveiling} proposed a refined model that better captures the QPO behavior, comprising two distinct components: a primary QPO centered 
around 5.8 Hz and a secondary QPO shoulder at approximately 6.3 Hz. Notably, 
at lower energy levels, the primary QPO dominates, whereas at higher energies, 
the secondary QPO shoulder becomes more prominent, effectively simulating the observed variation in the QPO frequency with energy. 
Additionally, the analysis of the rms and time lag spectra reveals notable discrepancies between the primary QPO and its 
shoulder counterpart. This discrepancy suggests that the shoulder phenomenon is not attributable to a mere drift in QPO frequency during observation but rather represents an independent component influencing the apparent energy dependence 
observed in the QPO frequency. However, as observed in the aforementioned sources and our results, the slope variations 
exhibit similar patterns, albeit with differences in the numerical values. This suggests that analogous phenomena should 
have a unified origin, thereby posing a challenge to the current QPO origin models. 

\subsection{QPO rms and phase lag}
In Figure~\ref{rmsmu}, a comprehensive comparison is presented, depicting the evolution of QPO rms with frequency. Below approximately 0.5 Hz, rms increases with frequency, followed by a decrease in rms across all energy bands. However, beyond approximately 1.5 Hz, LE QPO rms continues to decline, while the ME and HE bands remain at around 15$\%$ level, with a slow upward trend around 5 Hz. In the study of rms-frequency relationships, many results were analyzed specifically for QPOs in lower energy bands due to the limitations imposed by energy bands. 

In a study of GRS 1915+105, the transition frequency, where the rising trend shifts to a decline, is approximately 2 Hz \citep{yan2013statistical}.  The study of EXO 1846–031 using observations from the \textit{Insight}-HXMT similarly identifies a transition frequency of approximately 2 Hz, with the rms-frequency morphology in the low-energy (LE), medium-energy (ME), and high-energy (HE) bands resembling our findings \citep{liu2021timing}. In the study of MAXI J1803-298, similar results were observed, indicating a comparable transition frequency \citep{zhu2023timing}. \cite{shui2023tracing} present a systematic analysis of Type-C QPO observations of H1743-322. They employed the Lense-Thirring precession to fit the relationship between QPO rms and frequency during both rise and decay phases. Our results in Figure~\ref{rmsmu} are consistent with the behavior observed during the rise phases in their study. So this model should also be applicable to our data. However, the photon pile-up effect in the \textit{Insight}-HXMT LE spectra hinders the application of this model for fitting purposes.

In the QPO rms amplitude-energy relationship, the observed shapes in Figure~\ref{rmsenergy} is similar to that of many other sources, like MAXI J1803$-$298 \citep{zhu2023timing}, MAXI J1535$-$571 \citep{garg2022energy,kong2020joint,huang2018}, XTE J1550$-$564 \citep{li2013energy}, MAXI J1631$-$479 \citep{bu2021broadband}. \cite{you2018x} computed the fractional rms spectrum of the QPO within the framework of the Lense-Thirring precession model \citep{ingram2009low}, aiming to investigate energy-dependent variability during the state transition. Their computational findings indicate that at high inclinations, the rms spectrum increases with energy for energies below 10 keV, while beyond 10 keV, the rms spectrum gradually flattens. In our results, as clearly depicted in Figure~\ref{rmsenergy}, such a relationship is evident. 

Through analyzing a sample of 15 Galactic black hole binaries, \cite{vandeneijnden2017} uncovered a robust correlation between the phase lag of Type-C QPOs and the inclination angle: at lower frequencies, regardless of whether the source has a low or high inclination, the phase lag remains positive; as the frequency increases, sources with higher inclination gradually transition to soft lags, while those with lower inclination persist with hard lags. 
In Figure~\ref{phasefre}, we illustrate the relationship between QPO phase lag and frequency. At low frequencies, the phase lag maintains a hard lag, while around 1.5 Hz, the QPO phase lag begins to transition to a soft lag. 

A moderate inclination in this source is indicated by spectral fitting results 
 \citep{draghis2023preliminary,peng2024nicer}, and similarly, the inclination estimated from the averaged polarization degree \citep{veledina2023discovery} also suggests a moderate inclination. 
\cite{ingram2023tracking} investigated the evolution of the time lag of QPOs observed by NICER as a function of frequency for Swift J1727.8$-$1613  (see their Figure 6). Around 0.8 Hz, a hard lag is evident, which then transitions to a soft lag as the frequency gradually increases, becoming increasingly softer. This trend aligns with the results obtained from our calculations. Hence, Swift J1727.8$-$1613 should be a source that does not conform to the statistical results obtained by \cite{vandeneijnden2017}. \cite{WeiYu2024} provided temporal analysis results for Swift J1727.8$-$1613, indicating that based on the behavior of QPO rms and phase lag, this source is likely to have a high inclination, consistent with our findings. 

The inconsistency in inclination estimates may arise from the Lense-Thirring precession model not fully applying to Swift J1727.8$-$1613. \cite{zhao2024first} conducted an analysis of the polarimetric data of this source and performed a phase-resolved analysis of the QPO using the Hilbert-Huang transform technique. The photon index displayed a significant modulation relative to the QPO phase. However, the PD and PA exhibited no modulations in relation to the QPO phase, contrary to the anticipated behavior of the Lense-Thirring precession of the inner flow. \cite{ingram2023tracking} found that the variation in soft lag amplitude with spectral state deviates from the typical trend observed in other sources, suggesting that Swift J1727.8–1613 belongs to a previously underrepresented sub-population. 

\section{CONCLUSION}
\label{conclu}
We conducted an analysis of the Type-C QPOs phenomenon observed in Swift J1727.8$-$1613 using the \textit{Insight}-HXMT observations. The main results of the study can be summarized as follows: 

\begin{itemize}
    \item During the observation period, the flux in  HE continuously decreased, while the flux in the ME and  LE rapidly increased before decreasing again. However, two flare events occurred in the LE, during which the hardness ratio exhibited a noticeable decrease. The Type-C QPOs was observed almost throughout the entire observation period of \textit{Insight}-HXMT, with its frequency gradually increasing over time. During the flare-1 and flare-2 phases, there was a sudden increase of the QPO frequency. 
    
    \item The frequency of the QPOs exhibits a noticeable dependency 
    on energy. The observed phenomenon on the frequency-energy spectrum likely arises from a common underlying physical mechanism. Current theories are unable to fully explain it, necessitating further research. 

    \item The observed rms spectrum and phase lag spectrum of the Type-C QPOs resemble those of sources with high inclinations. However, analysis of the spectral fitting and polarization data suggests that this source has an inclination around 40 degrees. This contradiction may arise because this source belongs to a previously underrepresented sub-population. 

\end{itemize}

\begin{acknowledgments}
\section{acknowledgments}
 We are grateful to the referee for the useful suggestions to improve the manuscript. This work is supported by the National Key Research and Development Program of China (Grants No. 2021YFA0718503), the NSFC (12133007). This work has made use of data from the \textit{Insight-}HXMT mission, a project funded by the China National Space Administration (CNSA) and the Chinese Academy of Sciences (CAS).
  
\end{acknowledgments}

%







\newpage
\bibliography{sample631}{}

\begin{thebibliography}{}
\expandafter\ifx\csname natexlab\endcsname\relax\def\natexlab#1{#1}\fi
\providecommand{\url}[1]{\href{#1}{#1}}
\providecommand{\dodoi}[1]{doi:~\href{http://doi.org/#1}{\nolinkurl{#1}}}
\providecommand{\doeprint}[1]{\href{http://ascl.net/#1}{\nolinkurl{http://ascl.net/#1}}}
\providecommand{\doarXiv}[1]{\href{https://arxiv.org/abs/#1}{\nolinkurl{https://arxiv.org/abs/#1}}}

\bibitem[{Basak \& Zdziarski(2016)}]{basak2016spectral}
Basak, R., \& Zdziarski, A.~A. 2016, MNRAS, 458, 2199

\bibitem[{Belloni \& Hasinger(1990)}]{belloni1990atlas}
Belloni, T., \& Hasinger, G. 1990, A\&A, 230, 103

\bibitem[{Belloni {et~al.}(2016)Belloni, Motta, \& Bambi}]{belloni2016astrophysics}
Belloni, T., Motta, S., \& Bambi, C. 2016, Astrophysics and Space Science Library, Vol. 440, Astrophysics of Black Holes: From Fundamental Aspects to Latest Developments,  Springer-Verlag, Berlin

\bibitem[{Bollemeijer {et~al.}(2023)Bollemeijer, Uttley, Buisson, Homan, Altamirano, Gendreau, Arzoumanian, Strohmayer, \& Sanna}]{bollemeijer2023nicer}
Bollemeijer, N., Uttley, P., Buisson, D., {et~al.} 2023, ATel, 16247, 1

\bibitem[{Bu {et~al.}(2021)Bu, Zhang, Santangelo, Belloni, Zhang, Qu, Tao, Huang, Ma, Li, {et~al.}}]{bu2021broadband}
Bu, Q., Zhang, S., Santangelo, A., {et~al.} 2021, ApJ, 919, 92

\bibitem[{Bu {et~al.}(2015)Bu, Chen, Li, Qu, Belloni, \& Zhang}]{bu2015correlations}
Bu, Q.-c., Chen, L., Li, Z.-s., {et~al.} 2015, ApJ, 799, 2

\bibitem[{Cao {et~al.}(2020)Cao, Jiang, Meng, Zhang, Luo, Yang, Zhang, Gu, Sun, Liu, {et~al.}}]{cao2020medium}
Cao, X., Jiang, W., Meng, B., {et~al.} 2020, SCPMA, 63, 1

\bibitem[{Casella {et~al.}(2005)Casella, Belloni, \& Stella}]{casella2005abc}
Casella, P., Belloni, T., \& Stella, L. 2005, ApJ, 629, 403

\bibitem[{Castro-Tirado {et~al.}(2023)Castro-Tirado, Sanchez-Ramirez, Caballero-Garcia, Perez-Garcia, Fernandez-Garcia, Guziy, Hu, Blazek, Hermelo, Pinter, {et~al.}}]{castro2023optical}
Castro-Tirado, A., Sanchez-Ramirez, R., Caballero-Garcia, M., {et~al.} 2023, ATel, 16208, 1

\bibitem[{Chakrabarti {et~al.}(2008)Chakrabarti, Debnath, Nandi, \& Pal}]{chakrabarti2008evolution}
Chakrabarti, S.~K., Debnath, D., Nandi, A., \& Pal, P. 2008, \aap, 489, L41

\bibitem[{{Chakrabarti} \& {Molteni}(1993)}]{chakrabarti1993smoothed}
{Chakrabarti}, S.~K., \& {Molteni}, D. 1993, \apj, 417, 671

\bibitem[{Chatterjee {et~al.}(2024)Chatterjee, Mondal, Singh, \& Sugizaki}]{chatterjee2024insight}
Chatterjee, K., Mondal, S., Singh, C.~B., \& Sugizaki, M. 2024, arXiv preprint arXiv:2405.01498

\bibitem[{Chen {et~al.}(2020)Chen, Cui, Li, Wang, Xu, Lu, Wang, Chen, Han, Hu, {et~al.}}]{chen2020low}
Chen, Y., Cui, W., Li, W., {et~al.} 2020, SCPMA, 63, 1

\bibitem[{{Dichiara} {et~al.}(2023){Dichiara}, {Kennea}, {Page}, {Parsotan}, {Williams}, \& {Neil Gehrels Swift Observatory Team}}]{Kennea2023GCN}
{Dichiara}, S., {Kennea}, J.~A., {Page}, K.~L., {et~al.} 2023, GCN, 34542, 1

\bibitem[{Dovciak {et~al.}(2023{\natexlab{a}})Dovciak, Ratheesh, Tennant, \& Ma}]{dovciak2023ixpe}
Dovciak, M., Ratheesh, A., Tennant, A., \& Ma, G. 2023{\natexlab{a}}, ATel, 16237, 1

\bibitem[{Dovciak {et~al.}(2023{\natexlab{b}})Dovciak, Ratheesh, Tennant, \& Ma}]{dovciak2023ixpeb}
---. 2023{\natexlab{b}}, ATel, 16242, 1

\bibitem[{Draghis {et~al.}(2023)Draghis, Miller, Homan, Uttley, Bollemeijer, Steiner, Hare, Tombesi, Gendreau, Arzoumanian, {et~al.}}]{draghis2023preliminary}
Draghis, P.~A., Miller, J.~M., Homan, J., {et~al.} 2023, ATel, 16219, 1

\bibitem[{Garg {et~al.}(2022)Garg, Misra, \& Sen}]{garg2022energy}
Garg, A., Misra, R., \& Sen, S. 2022, MNRAS, 514, 3285

\bibitem[{Homan \& Belloni(2005)}]{homan2005evolution}
Homan, J., \& Belloni, T. 2005, Astrophysics and Space Science: From X-Ray Binaries to Quasars: Black Holes on all Mass Scales, 107

\bibitem[{Huang {et~al.}(2018)Huang, Qu, Zhang, Bu, Chen, Tao, Zhang, Lu, Li, Song, {et~al.}}]{huang2018}
Huang, Y., Qu, J., Zhang, S., {et~al.} 2018, ApJ, 866, 122

\bibitem[{Huppenkothen {et~al.}(2019)Huppenkothen, Bachetti, Stevens, Migliari, Balm, Hammad, Khan, Mishra, Rashid, Sharma, {et~al.}}]{huppenkothen2019stingray}
Huppenkothen, D., Bachetti, M., Stevens, A.~L., {et~al.} 2019, ApJ, 881, 39

\bibitem[{Ingram \& Done(2011)}]{ingram2011physical}
Ingram, A., \& Done, C. 2011, MNRAS, 415, 2323

\bibitem[{Ingram {et~al.}(2009)Ingram, Done, \& Fragile}]{ingram2009low}
Ingram, A., Done, C., \& Fragile, P.~C. 2009, MNRAS: Letters, 397, L101

\bibitem[{Ingram {et~al.}(2016)Ingram, van~der Klis, Middleton, Done, Altamirano, Heil, Uttley, \& Axelsson}]{ingram2016quasi}
Ingram, A., van~der Klis, M., Middleton, M., {et~al.} 2016, MNRAS, 461, 1967

\bibitem[{Ingram {et~al.}(2023)Ingram, Bollemeijer, Veledina, Dovciak, Poutanen, Egron, Russell, Trushkin, Negro, Ratheesh, {et~al.}}]{ingram2023tracking}
Ingram, A., Bollemeijer, N., Veledina, A., {et~al.} 2023, arXiv preprint arXiv:2311.05497

\bibitem[{Ingram(2019)}]{ingram2019}
Ingram, A.~R. 2019, NewAR, 29

\bibitem[{Karpouzas {et~al.}(2021)Karpouzas, M{\'e}ndez, Garc{\'\i}a, Zhang, Altamirano, Belloni, \& Zhang}]{karpouzas2021variable}
Karpouzas, K., M{\'e}ndez, M., Garc{\'\i}a, F., {et~al.} 2021, MNRAS, 503, 5522

\bibitem[{Katoch {et~al.}(2023)Katoch, Antia, Nandi, \& Shah}]{katoch2023detection}
Katoch, T., Antia, H., Nandi, A., \& Shah, P. 2023, ATel, 16235, 1

\bibitem[{Kong {et~al.}(2020)Kong, Zhang, Chen, Ji, Zhang, Yang, Tao, Ma, Qu, Lu, {et~al.}}]{kong2020joint}
Kong, L., Zhang, S., Chen, Y., {et~al.} 2020, JHEAp, 25, 29

\bibitem[{Li {et~al.}(2013{\natexlab{a}})Li, Qu, Song, Ding, \& Zhang}]{li2013energy}
Li, Z., Qu, J., Song, L., Ding, G., \& Zhang, C. 2013{\natexlab{a}}, MNRAS, 428, 1704

\bibitem[{Li {et~al.}(2013{\natexlab{b}})Li, Zhang, Qu, Gao, Zhao, Huang, \& Song}]{li2013energyH1743}
Li, Z., Zhang, S., Qu, J., {et~al.} 2013{\natexlab{b}}, MNRAS, 433, 412

\bibitem[{Liu {et~al.}(2020)Liu, Zhang, Li, Lu, Chang, Li, Zhang, Jin, Yu, Zhang, {et~al.}}]{liu2020High}
Liu, C., Zhang, Y., Li, X., {et~al.} 2020, SCPMA, 63, 1

\bibitem[{Liu {et~al.}(2021)Liu, Huang, Xiao, Bu, Qu, Zhang, Zhang, Jia, Lu, Ma, {et~al.}}]{liu2021timing}
Liu, H.-X., Huang, Y., Xiao, G.-C., {et~al.} 2021, RAA, 21, 070

\bibitem[{Ma {et~al.}(2021)Ma, Tao, Zhang, Zhang, Bu, Ge, Chen, Qu, Zhang, Lu, {et~al.}}]{ma2021discovery}
Ma, X., Tao, L., Zhang, S.-N., {et~al.} 2021, NatAs, 5, 94

\bibitem[{Ma {et~al.}(2023)Ma, Zhang, Tao, Bu, Qu, Zhang, Zhou, Huang, Jia, Song, {et~al.}}]{ma2023detailed}
Ma, X., Zhang, L., Tao, L., {et~al.} 2023, ApJ, 948, 116

\bibitem[{Marcel \& Neilsen(2021)}]{marcel2021can}
Marcel, G., \& Neilsen, J. 2021, ApJ, 906, 106

\bibitem[{M{\'e}ndez {et~al.}(2024)M{\'e}ndez, Peirano, Garc{\'\i}a, Belloni, Altamirano, \& Alabarta}]{mendez2024unveiling}
M{\'e}ndez, M., Peirano, V., Garc{\'\i}a, F., {et~al.} 2024, MNRAS, 527, 9405

\bibitem[{Mereminskiy {et~al.}(2023)Mereminskiy, Lutovinov, Molkov, Krivonos, Semena, Sazonov, Tkachenko, \& Sunyaev}]{mereminskiy2023hard}
Mereminskiy, I., Lutovinov, A., Molkov, S., {et~al.} 2023, arXiv preprint arXiv:2310.06697

\bibitem[{Miller-Jones {et~al.}(2023)Miller-Jones, Sivakoff, Bahramian, \& Russell}]{miller2023vla}
Miller-Jones, J., Sivakoff, G., Bahramian, A., \& Russell, T. 2023, ATel, 16211, 1

\bibitem[{Miyamoto {et~al.}(1991)Miyamoto, Kimura, Kitamoto, Dotani, \& Ebisawa}]{miyamoto1991x}
Miyamoto, S., Kimura, K., Kitamoto, S., Dotani, T., \& Ebisawa, K. 1991, ApJ, 383, 784

\bibitem[{{Molteni} {et~al.}(1996){Molteni}, {Sponholz}, \& {Chakrabarti}}]{Molteni1996}
{Molteni}, D., {Sponholz}, H., \& {Chakrabarti}, S.~K. 1996, \apj, 457, 805

\bibitem[{Motta {et~al.}(2015)Motta, Casella, Henze, Mu{\~n}oz-Darias, Sanna, Fender, \& Belloni}]{motta2015geometrical}
Motta, S., Casella, P., Henze, M., {et~al.} 2015, MNRAS, 447, 2059

\bibitem[{Motta(2016)}]{motta2016quasi}
Motta, S.~E. 2016, AN, 337, 398

\bibitem[{Nandi {et~al.}(2024)Nandi, Das, Majumder, Katoch, Antia, \& Shah}]{nandi2024discovery}
Nandi, A., Das, S., Majumder, S., {et~al.} 2024, arXiv preprint arXiv:2404.17160

\bibitem[{Nathan {et~al.}(2022)Nathan, Ingram, Homan, Huppenkothen, Uttley, van~der Klis, Motta, Altamirano, \& Middleton}]{nathan2022phase}
Nathan, E., Ingram, A., Homan, J., {et~al.} 2022, MNRAS, 511, 255

\bibitem[{Negoro {et~al.}(2023)Negoro, Serino, Nakajima, Kobayashi, Tanaka, Soejima, Kudo, Mihara, Kawamuro, Yamada, {et~al.}}]{negoro2023maxi}
Negoro, H., Serino, M., Nakajima, M., {et~al.} 2023, ATel, 16205, 1

\bibitem[{O'Connor {et~al.}(2023)O'Connor, Hare, Younes, Gendreau, Arzoumanian, \& Ferrara}]{o2023nicer}
O'Connor, B., Hare, J., Younes, G., {et~al.} 2023, GCN, 34549, 1

\bibitem[{{Page} {et~al.}(2023){Page}, {Dichiara}, {Gropp}, {Krimm}, {Parsotan}, {Williams}, \& {Neil Gehrels Swift Observatory Team}}]{Page2023}
{Page}, K.~L., {Dichiara}, S., {Gropp}, J.~D., {et~al.} 2023, GCN, 34537, 1

\bibitem[{Palmer \& Parsotan(2023)}]{palmer2023swift}
Palmer, D.~M., \& Parsotan, T.~M. 2023, ATel, 16215, 1

\bibitem[{Peng {et~al.}(2024)Peng, Zhang, Shui, Zhang, Kong, Chen, Wang, Ji, Qu, Tao, {et~al.}}]{peng2024nicer}
Peng, J.-Q., Zhang, S., Shui, Q.-C., {et~al.} 2024, ApJL, 960, L17

\bibitem[{Qu {et~al.}(2010)Qu, Lu, Lu, Song, Zhang, Ding, \& Wang}]{qu2010energy}
Qu, J., Lu, F., Lu, Y., {et~al.} 2010, ApJ, 710, 836

\bibitem[{Remillard \& McClintock(2006)}]{remillard2006}
Remillard, R.~A., \& McClintock, J.~E. 2006, ARA\&A, 44, 49

\bibitem[{S{\'a}nchez {et~al.}(2024)S{\'a}nchez, Mu{\~n}oz-Darias, Padilla, Casares, \& Torres}]{sanchez2024evidence}
S{\'a}nchez, D.~M., Mu{\~n}oz-Darias, T., Padilla, M.~A., Casares, J., \& Torres, M. 2024, \aap, 682, L1

\bibitem[{Schnittman {et~al.}(2006)Schnittman, Homan, \& Miller}]{schnittman2006precessing}
Schnittman, J.~D., Homan, J., \& Miller, J.~M. 2006, ApJ, 642, 420

\bibitem[{Sharma {et~al.}(2018)Sharma, Jaleel, Jain, Pandey, Paul, \& Dutta}]{sharma2018spectral}
Sharma, R., Jaleel, A., Jain, C., {et~al.} 2018, MNRAS, 481, 5560

\bibitem[{Shui {et~al.}(2023)Shui, Zhang, Chen, Zhang, Kong, Wang, Ji, Yin, Qu, Tao, {et~al.}}]{shui2023tracing}
Shui, Q.~C., Zhang, S., Chen, Y.~P., {et~al.} 2023, ApJ, 943, 165

\bibitem[{Stella \& Vietri(1997)}]{stella1997lense}
Stella, L., \& Vietri, M. 1997, ApJ, 492, L59

\bibitem[{Stella {et~al.}(1999)Stella, Vietri, \& Morsink}]{stella1999correlations}
Stella, L., Vietri, M., \& Morsink, S.~M. 1999, ApJ, 524, L63

\bibitem[{Sunyaev {et~al.}(2023)Sunyaev, Mereminskiy, Molkov, Semena, Arefiev, Krivonos, Levin, Lutovinov, Shtykovsky, \& Tkachenko}]{sunyaev2023integral}
Sunyaev, R., Mereminskiy, I., Molkov, S., {et~al.} 2023, ATel, 16217, 1

\bibitem[{{Tagger} \& {Pellat}(1999)}]{Tagger1999}
{Tagger}, M., \& {Pellat}, R. 1999, \aap, 349, 1003

\bibitem[{van~den Eijnden {et~al.}(2016)van~den Eijnden, Ingram, \& Uttley}]{van2016probing}
van~den Eijnden, J., Ingram, A., \& Uttley, P. 2016, MNRAS, 458, 3655

\bibitem[{Van~den Eijnden {et~al.}(2016)Van~den Eijnden, Ingram, Uttley, Motta, Belloni, \& Gardenier}]{van2016inclination}
Van~den Eijnden, J., Ingram, A., Uttley, P., {et~al.} 2016, MNRAS, stw2634

\bibitem[{van~den Eijnden {et~al.}(2017)van~den Eijnden, Ingram, Uttley, Motta, Belloni, \& Gardenier}]{vandeneijnden2017}
van~den Eijnden, J., Ingram, A., Uttley, P., {et~al.} 2017, MNRAS, 464, 2643

\bibitem[{Varniere \& Tagger(2002)}]{varniere2002accretion}
Varniere, P., \& Tagger, M. 2002, \aap, 394, 329

\bibitem[{Veledina {et~al.}(2023)Veledina, Muleri, Dov{\v{c}}iak, Poutanen, Ratheesh, Capitanio, Matt, Soffitta, Tennant, Negro, {et~al.}}]{veledina2023discovery}
Veledina, A., Muleri, F., Dov{\v{c}}iak, M., {et~al.} 2023, ApJL, 958, L16

\bibitem[{Yan {et~al.}(2013)Yan, Ding, Wang, Qu, \& Song}]{yan2013statistical}
Yan, S.-P., Ding, G.-Q., Wang, N., Qu, J.-L., \& Song, L.-M. 2013, MNRAS, 434, 59

\bibitem[{Yan {et~al.}(2012)Yan, Qu, Ding, Han, Song, Zhang, Yin, Zhang, \& Wang}]{yan2012systematic}
Yan, S.-P., Qu, J.-L., Ding, G.-Q., {et~al.} 2012, Ap\&SS, 337, 137

\bibitem[{You {et~al.}(2018)You, Bursa, \& {\.Z}ycki}]{you2018x}
You, B., Bursa, M., \& {\.Z}ycki, P.~T. 2018, ApJ, 858, 82

\bibitem[{Yu {et~al.}(2024)Yu, Bu, Zhang, Liu, Zhang, Ducci, Tao, Santangelo, Doroshenko, Huang, Yang, \& Qu}]{WeiYu2024}
Yu, W., Bu, Q.-C., Zhang, S.-N., {et~al.} 2024, arXiv preprint arXiv:2403.13127

\bibitem[{Zhang {et~al.}(2017)Zhang, Wang, M{\'e}ndez, Chen, Qu, Altamirano, \& Belloni}]{zhang2017evolution}
Zhang, L., Wang, Y., M{\'e}ndez, M., {et~al.} 2017, ApJ, 845, 143

\bibitem[{Zhao {et~al.}(2024)Zhao, Tao, Li, Zhang, Feng, Ge, Ji, Wang, Huang, Ma, {et~al.}}]{zhao2024first}
Zhao, Q.-C., Tao, L., Li, H.-C., {et~al.} 2024, ApJL, 961, L42

\bibitem[{Zhu {et~al.}(2023)Zhu, Chen, \& Wang}]{zhu2023timing}
Zhu, H., Chen, X., \& Wang, W. 2023, MNRAS, 523, 4394

\end{thebibliography}
\bibliographystyle{aasjournal}



\end{document}